\documentclass{PoS}
\usepackage{graphicx}
\usepackage{wrapfig}
\usepackage{wrapfig,booktabs}
\usepackage{color}
\usepackage{multirow}
\usepackage{array}
\usepackage{float}
\usepackage{amsmath,amssymb,amsfonts,amstext,amsthm}
\newcommand{\bea}{\begin{eqnarray}}
\newcommand{\eea}{\end{eqnarray}}
\newcommand{\BAN}{\begin{eqnarray*}}
\newcommand{\EAN}{\end{eqnarray*}}
\def\Id{\mbox{1\hspace{-1.2mm}I} }

\newcommand{\p}{\partial}
\newcommand{\pslash}{p\kern-1ex /}
\newcommand{\lslash}{l\kern-1ex /}
\newcommand{\kslash}{k\kern-1ex /}
\newcommand{\dslash}{\p\kern-1.2ex /}
\newcommand{\Dslash}{{\cal D}\kern-1.5ex /}

\newcommand{\tr}{{\rm tr}}

\def\u{{\bf u}}
\def\d{{\bf d}}
\def\s{{\bf s}}
\def\c{{\bf c}}

\def\q{{\bf q}}

\def\ubar{\bar{\bf u}}

\def\sbar{\bar{\bf s}}
\def\cbar{\bar{\bf c}}

\def\qbar{\bar{\bf q}}

\title{First study of $N_f=2+1+1$ lattice QCD with physical domain-wall quarks}

\ShortTitle{$N_f=2+1+1$ lattice QCD with physcial domain-wall quarks}

\author{\speaker{Ting-Wai Chiu}\thanks{This work is supported by the Ministry of Science and Technology
(Nos.~107-2119-M-003-008, 108-2112-M-003-005)} \ (TWQCD collaboration) \\
        Physics Department, National Taiwan Normal University, Taipei, Taiwan 11677, R.O.C.\\
        Institute of Physics, Academia Sinica, Taipei, Taiwan 11529, R.O.C.\\
        Physics Department, National Taiwan University, Taipei, Taiwan 10617, R.O.C.\\
        E-mail: \email{twchiu@phys.ntu.edu.tw}
}

\abstract{
Using 10-16 units of Nvidia DGX-1, we have generated the first gauge ensemble for $N_f=2+1+1$ lattice QCD 
with physical $(\u/\d, \s, \c)$ domain-wall quarks, on the $ 64^4 $ lattice with 
lattice spacing $a \sim 0.064 $~fm ($ L > 4 $ fm, and $ M_\pi L > 3 $). 
The salient feature of this gauge ensemble is that the chiral symmetry is preserved to a high precision 
and all topological sectors are sampled ergodically.  
In this paper, we present the first results of the topological susceptibility and 
the ground-state hadron mass spectra. 
}

\FullConference{37th International Symposium on Lattice Field Theory - Lattice2019\\
		16-22 June 2019\\
		Wuhan, China}

\begin{document}

\section{Introduction}

The holy grail of lattice QCD is to simulate QCD at the physical point, with
sufficiently large volume and fine lattice spacing, then to extract physics from these gauge ensembles.
Nevertheless, the characteristics of a gauge ensemble depends on the lattice action used for the simulation, 
which in turn has significant impacts on the physics outcome. 

Since quarks are Dirac fermions, 
they possess the chiral symmetry in the massless limit. At the zero temperature, 
the chiral symmetry $ SU_L(N_f) \times SU_R(N_f) $ of $ N_f $ massless quarks 
is spontaneously broken to $ SU_V(N_f) $, due to the strong interaction between quarks and gluons
in the QCD vacuum (with non-trivial topology).
The quark condensate $\left<\qbar\q\right>$ constitutes the origin of hadron masses, and 
resolves the puzzles such as ``why the mass of a proton is much heavier than the sum 
of the bare quark masses of its constituents". 
To investigate the spontaneous chiral symmetry breaking
as well as the hadron physics (e.g., the mass spectrum) from the first principles of QCD,
it requires nonperturbative methods. 
So far, lattice QCD is the most promising approach,
discretizing the continuum space-time on a 4-dimensional lattice,
and computing physical observables by Monte Carlo simulation.

However, in lattice QCD, formulating lattice fermion
with exact chiral symmetry at finite lattice spacing is rather nontrivial. This
is realized through domain-wall fermion (DWF) on the 5-dimensional lattice \cite{Kaplan:1992bt}
and the overlap-Dirac fermion on the 4-dimensional lattice \cite{Neuberger:1997fp,Narayanan:1994gw}.
Nevertheless, the computational requirement for lattice QCD with domain-wall quarks on a 
5-dimensional lattice is 10-100 times more than their counterparts with traditional 
lattice fermions (e.g., Wilson, staggered, and their variants). This is one of the reasons 
why there are only 3 lattice QCD groups worldwide (RBC/UKQCD, JLQCD, TWQCD)  
using DWF for large-scale lattice QCD simulations.  

Since 2018, TWQCD has been performing the hybrid Monte Carlo (HMC) simulation of $N_f=2+1+1$ lattice QCD 
at the physical point \cite{Chiu:2018qcp} with the optimal domain-wall fermion 
\cite{Chiu:2002ir}, 
on the $ 64^4 $ lattice with the extent $ N_s = 16 $ in the fifth dimension, following our first  
$N_f=2+1+1$ simulation on the $32^3 \times 64 $ lattice 
\cite{Chen:2017kxr}.
It is interesting to point out that the entire simulation on the $64^4 \times 16$ lattice 
can be fitted into one unit of Nvidia DGX-1,  
which consists of eight V100 GPUs interconnected by the NVLink, 
with the total device memory $8 \times 16~\mbox{GB} = 128$~GB.
In general, to simulate $N_f = 2+1+1$ lattice QCD on the $ 64^4 $ lattice 
with any lattice Dirac operator $ D $ requires memory much larger than 128~GB, 
since each one-flavor pseudofermion action is expressed as the rational approximation 
of $ \Phi^\dagger (D^\dagger D)^{-1/2} \Phi $, requiring a large number 
of long vectors (proportional to the number of poles in the rational approximation) 
in computing the pseudofermion force in the molecular dynamics. 
However, for domain-wall fermion, one can use the exact one-flavor pseudofermion action (EOFA) 
with a positive-definite and Hermitian Dirac operator \cite{Chen:2014hyy}, thus avoiding the  
rational approximation and saving a large amount of device memory.  
Moreover, using EOFA 
also enhances the HMC efficiency significantly. 

The thermalization is performed with one unit of Nvidia DGX-1, running for $\sim 8$ months. 
Then the thermalized configurations are distributed to 10-16 units of Nvidia DGX-1,  
each running an independent HMC simulation for $\sim 6-10$ months, resulting a total 
of $\sim 2050$ HMC trajectories. 
By sampling one configuration every 5 trajectories in each independent HMC simulation, 
we obtain $ 400 $ gauge configurations.  
The lattice setup and simulation parameters have been given in Ref. \cite{Chiu:2018qcp}. 
Now the lattice spacing is updated to 
$ a^{-1} = 3.188 \pm 0.018~{\text{GeV}} $ for 400 gauge configurations.
\begin{wraptable}{r}{0.55\textwidth}
\caption{The residual masses of $ \u/\d $, $\s $, and $ \c $ quarks.}
\begin{tabular}{cccc} \\\toprule
quark & $m_q a $ & $ m_{res} a $ & $ m_{res}$~[MeV]   \\\midrule
$\u/\d$ & 0.00125  & $ 5.77(17) \times 10^{-5} $ & 0.178(5)   \\ 
$\s$    & 0.040    & $ 1.45(12) \times 10^{-5} $ & 0.045(4)   \\
$\c$    & 0.550    & $ 0.21(4) \times 10^{-5} $ & 0.006(1)   \\  \bottomrule
\end{tabular}
\label{tab:mres}
\end{wraptable} 
Also, the residual masses of $(\u/\d, \s, \c)$ quarks are updated in Table \ref{tab:mres} 
for 400 configurations. 
For $ \u/\d $ quark, the residual mass is $\sim 4.6$\% of its bare mass, amounting to $0.178(5) $~MeV.
For $ \s $ and $ \c $ quarks, the residual masses are even smaller, 
$ 0.045(4)$~MeV, and $0.006(1) $~MeV respectively. 
This demonstrates that the optimal DWF can preserve the chiral symmetry 
to a high precision, for both light and heavy quarks.   
In the following, we present the first results of the topological susceptibility
and the hadron mass spectra.

\section{Topological Susceptibility}

The vacuum of QCD has a non-trivial topological structure.
The topological fluctuations of the QCD vacuum can be measured in terms of the moments of
the topological charge $ \langle Q_t^{2p} \rangle, p = 1, 2, \cdots $,  
where $ Q_t $ is the (integer-valued) topological charge of the gauge field,
\bea
\label{eq:Qt}
Q_t=\frac{\epsilon_{\mu\nu\lambda\sigma}}{32 \pi^2} \int d^4 x \ \tr[ F_{\mu\nu}(x) F_{\lambda\sigma}(x)],
\eea
and $ F_{\mu\nu} = g T^a F_{\mu\nu}^a$ is the matrix-valued field tensor, with the normalization
$ \tr (T^a T^b) = \delta_{ab}/2 $. 
Among all moments, the topological susceptibility $ \chi_t $
\bea
\label{eq:chit}
\chi_t = \frac{ \langle Q_t^2 \rangle }{\Omega}, \hspace{4mm} \Omega = \text{4-dimensional volume},
\eea
is the most crucial one, which plays the important role in breaking the $ U_A(1) $ symmetry,
and resolves the puzzle why the flavor-singlet $ \eta'$ is much
heavier than other non-singlet (approximate) Goldstone bosons. In general, it can be shown that 
the topological susceptibility and the quark condensates are closely related, which in turn implies that 
{\it a gauge ensemble without the proper topological susceptibility cannot give 
the correct hardon mass spectrum (or any physical observable) from the first principles of QCD.}  

In lattice QCD with exact chiral symmetry, 
the index of the massless overlap-Dirac operator 
is equal to $Q_t$, satisfying the Atiyah-Singer index theorem,  
$ \text{index}(D_{\text{ov}}) = Q_t $. 
However, to project the zero modes of the massless overlap-Dirac operator for 
the $ 64^4 $ lattice is prohibitively expensive.
On the other hand, the clover topological charge  
$ Q_{\text{clover}} = \sum_x \epsilon_{\mu\nu\lambda\sigma} 
\tr[ F_{\mu\nu}(x) F_{\lambda\sigma}(x) ]/(32 \pi^2) $
is not reliable [where the matrix-valued field tensor $ F_{\mu\nu}(x) $ is obtained from 
the four plaquettes surrounding $ x $ on the ($\hat\mu,\hat\nu$) plane],    
unless the gauge configuration is sufficiently smooth. 
Nevertheless, the smoothness of a gauge configuration 
can be attained by the Wilson flow \cite{Narayanan:2006rf,Luscher:2010iy}, 
which is a continuous-smearing process to average gauge field 
over a spherical region of root-mean-square radius $ R_{rms} = \sqrt{8 t} $, where $ t $ is 
the flow-time. 
As $t$ gets larger, the gauge configuration becomes smoother, and 
$ Q_{\text{clover}}(t) $ converges to its nearest integer, $ \text{round}[Q_{\text{clover}}(t)] $, 
which hardly changes for $ t/a^2 \gg 1 $. 
Consequently, the $ \chi_t(t) $ computed with $ Q_{\text{clover}}(t) $
behaves like a constant for $t/a^2 \gg 1 $.  
In other words, applying the Wilson flow to an ensemble of gauge configurations
for a sufficiently long flow-time can let them fall into topological sectors, 
similar to the gauge fields in the continuum theory. 
Moreover, it has been demonstrated that the asymptotic value of 
$ \chi_t $ computed with $ Q_{\text{clover}} $ is in good agreement with that 
computed with the index of overlap-Dirac operator at $t=0$ \cite{Chiu:2019xry}. 
Thus the topological susceptibility 
in lattice QCD with exact chiral symmetry 
can be obtained from the asymptotic value of 
$ \chi_t $ computed with $ Q_{\text{clover}} $ in the Wilson flow.
 
\begin{figure}[!ht]
\begin{center}
\begin{tabular}{@{}cccc@{}}
\includegraphics*[height=6cm,width=7cm,clip=true]{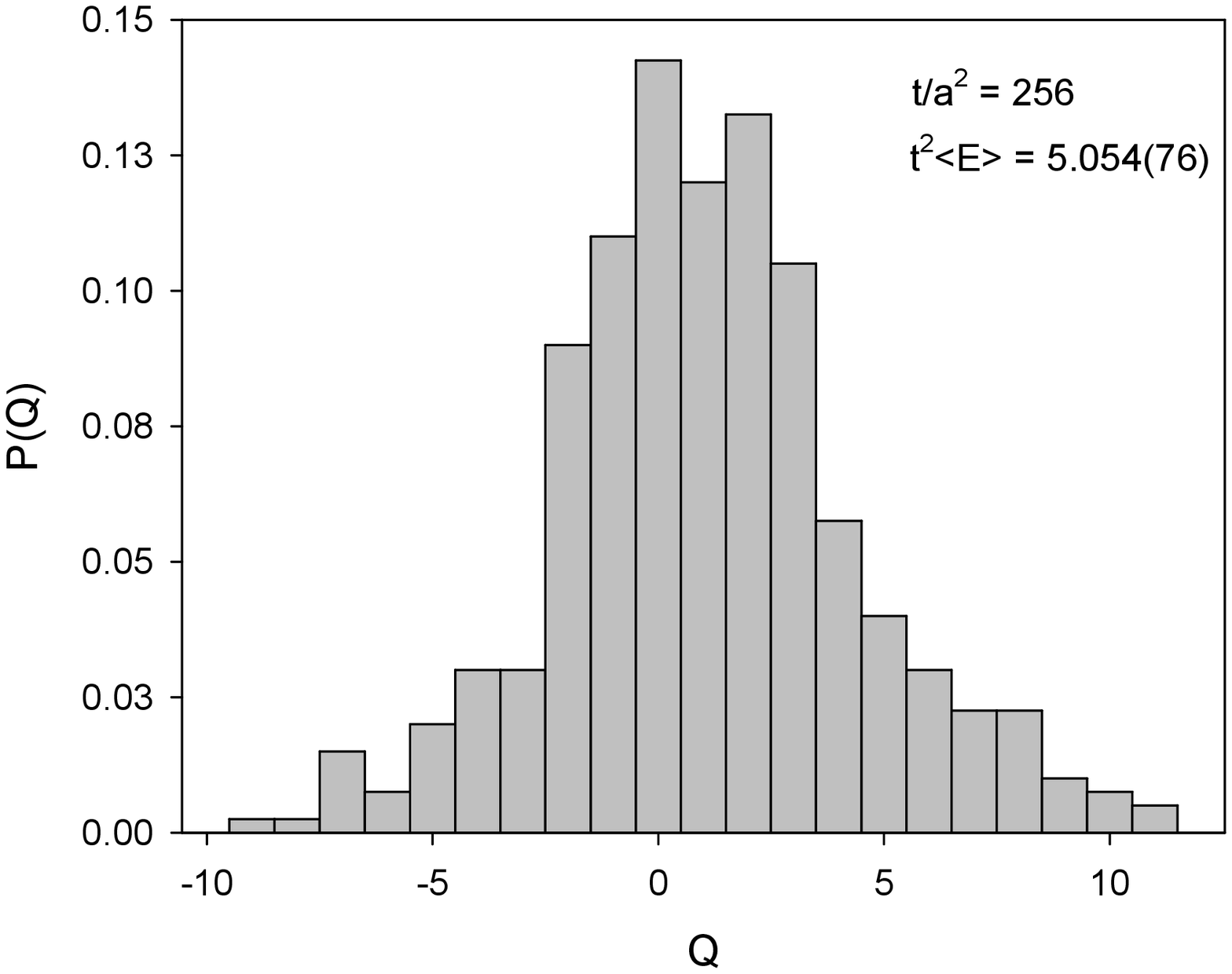}
&
\includegraphics*[height=6cm,width=7cm,clip=true]{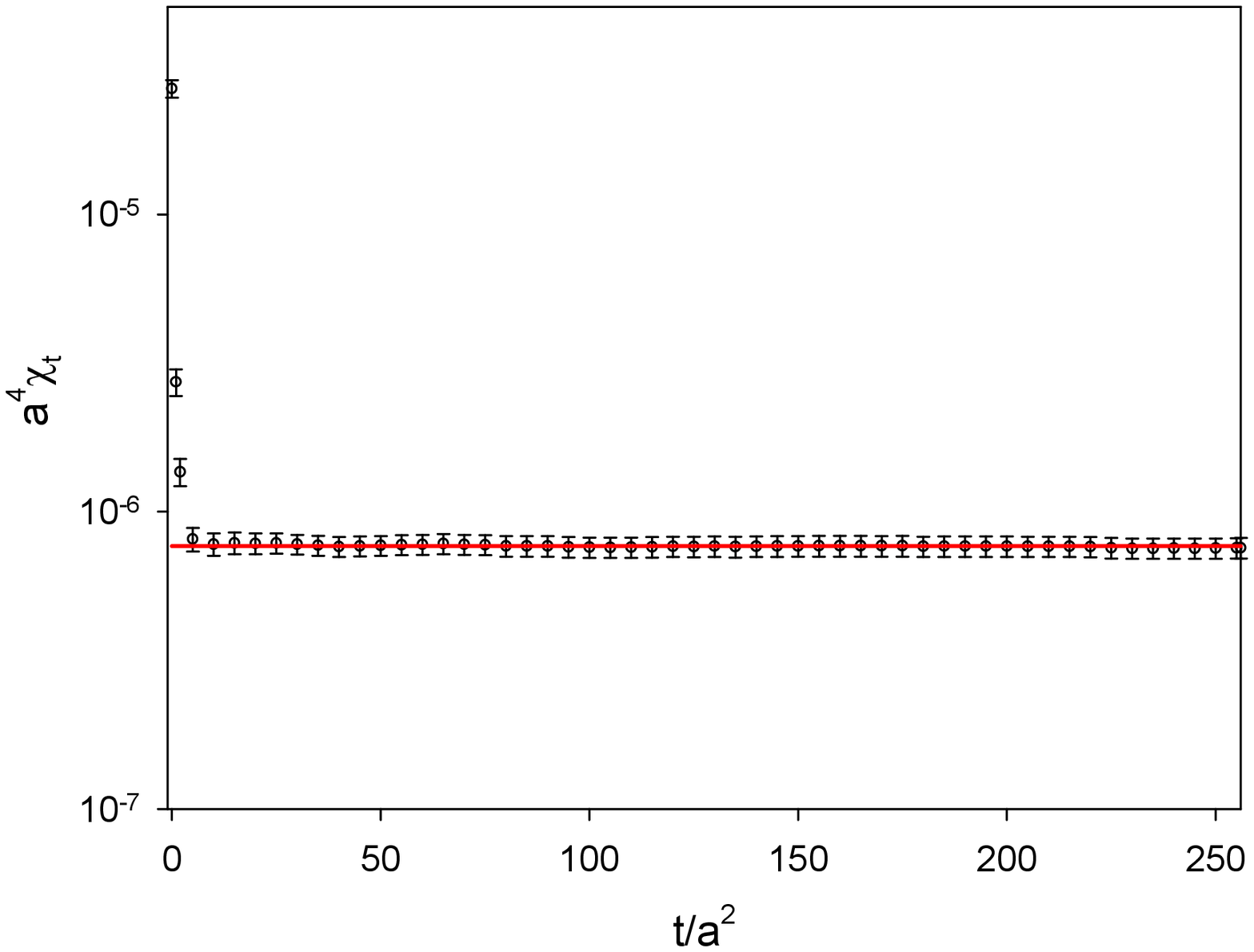}
\\
\end{tabular}
\caption{
(left panel) The histogram of the probability distribution of $Q_{\text{clover}}$ at $t/a^2 = 256 $.
(right panel) The topological susceptibility computed with $Q_{\text{clover}}(t)$ 
versus the flow-time $ t/a^2 $. The horizontal line is the constant fit for $ 30 \le t/a^2 \le 256 $.}
\label{fig:Q_clover_hist_chit}
\end{center}
\end{figure}

In this study, the flow equation is numerically integrated from $ t = 0 $ to 256 
with $ \Delta t = 0.01 $. In Fig.~\ref{fig:Q_clover_hist_chit}, 
the histogram of the probability distribution of $ Q_{\text{clover}}(t) $ 
at $ t/a^2 = 256 $ is plotted on the left panel, while the topological susceptibility 
computed with $Q_{\text{clover}}(t)$ versus the flow-time $ t/a^2 $ is plotted on the right panel. 
Evidently, the $ \chi_t $ becomes almost a constant for $ t/a^2 > 10 $. 
Fitting $ \chi_t $ to a constant for $ 30 \le t/a^2 \le 256 $ gives 
\bea
\chi_t a^4 = (7.66 \pm 0.42) \times 10^{-7}
\eea 
with $ \chi^2/\text{d.o.f.} = 0.11 $.  
The systematic error can be estimated by changing the range of $t$ for fitting as well as  
by replacing $Q_{\text{clover}}(t)$ with its nearest integer $ \text{round}[Q_{\text{clover}}(t)] $ 
in computing $ \chi_t $. The final result of $ \chi_t $ in the energy units is 
\bea
\label{eq:chit14_MeV}
\chi_t^{1/4} = ( 94.29 \pm 1.30 \pm 0.78 )~\text{MeV}, 
\eea
where the first/second uncertainty is the statistical/systematic one.

\section{Hardon Mass Spectrum}

One of the main objectives of lattice QCD is to extract the mass spectrum of QCD nonperturbatively 
from the first principles, and to compare it with the experimental data.  
%
%
To this end, the first step is to extract the ground-state hadron mass spectra  
from the time-correlation functions of quark-antiquark meson interpolators 
and 3-quark baryon interpolators respectively, and to check whether the theoretical results 
are compatible with the experimental mass spectra. 
If a theoretical state can be identified with an experimental state 
by the same $J^{P(C)}$ and the proximity of mass,  
then we can infer that this hadron state behaves like the 
conventional quark-antiquark meson or 3-quark baryon.  
On the other hand, if the mass of a theoretical state is incompatible 
with any experimental state with the same $J^{P(C)}$, then there could be two possibilities. 
It could be a state to be observed in the future experiments, thus serves as 
a prediction of lattice QCD. Another possibility is that the targeted hadron
is an exotic state. Thus the conventional quark-antiquark or the 3-quark interpolator
cannot overlap with all components of this exotic hadron
and gives a theoretical mass different from the experimental value.
In this case, further theoretical/experimental studies are required to clarify 
the nature of this exotic hadron state. 
Theoretically, for the exotic meson state, it requires to analyze 
the correlation matrix of interpolators consisting of quark-antiquark, meson-meson, diquark-antidiquark, 
and quark-antiquark-gluon operators; while for the exotic baryon state, 
to study the correlation matrix of interpolators consisting of 3-quark, meson-baryon,   
and diquark-diquark-antiquark operators.  
   
In the following, the time-correlation functions of local operators are measured with 
point-to-point quark propagators computed with the same
parameters ($N_s = 16 $, $m_0 = 1.3 $, $ \lambda_{max}/\lambda_{min} = 6.20/0.05 $)
and masses ($ m_{u/d} a = 0.00125, m_s a = 0.04, m_c a = 0.55 $) of the sea quarks, 
where $m_{u/d}$, $m_s$ and $m_c$ are fixed by the masses of $ \pi^{\pm}(140) $, 
$ \phi(1020) $ and $ J/\psi(3097) $ respectively. Then the masses of any other hadrons
containing $\u, \d, \s $ and $ \c $ quarks are predictions from the first principles of QCD. 
%

\begin{figure}[!ht]
\begin{center}
\includegraphics*[height=7.7cm,width=10.5cm,clip=true]{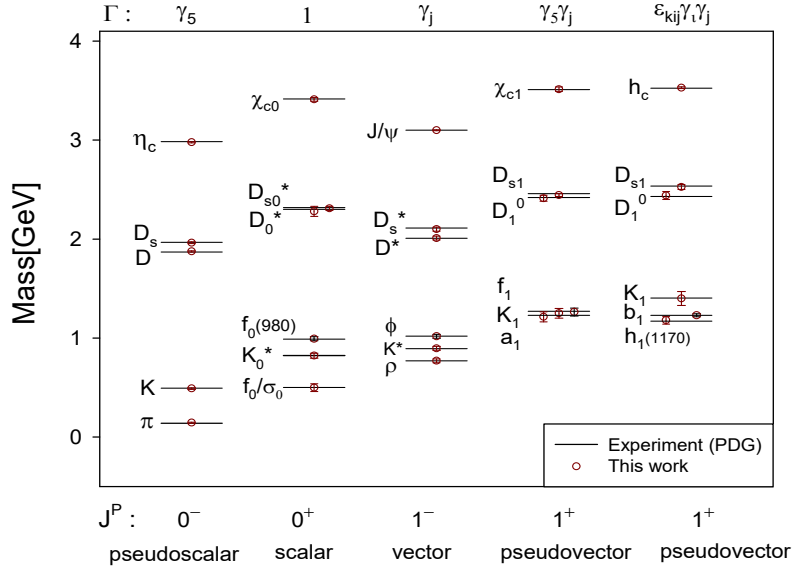}
\caption{
The ground-state masses of the quark-antiquark mesons with flavor contents   
$ \ubar\d $, $ \ubar \s $, $ \ubar \c $, $ \sbar \s $, 
$ \sbar \c $, and $ \cbar \c $, versus the experimental states \cite{Tanabashi:2018oca}.}
\label{fig:meson}
\end{center}
\end{figure}

\subsection{Meson mass spectrum}

The meson interpolators in this study are: 
$ \ubar \Gamma \d $, $ \ubar \Gamma \s $, $ \ubar \Gamma \c $, $ \sbar \Gamma \s $, 
$ \sbar \Gamma \c $, and $ \cbar \Gamma \c $, where 
$\Gamma=\{\Id,\gamma_5,\gamma_i,\gamma_5\gamma_i, \epsilon_{ijk}\gamma_j\gamma_k \} $,  
corresponding to scalar ($S$), pseudoscalar ($P$), vector ($V$), pseudovector ($A$),
and pseudovector ($T$) respectively. 
Note that $ \qbar \gamma_5\gamma_i \q $ transforms like $ J^{PC} = 1^{++} $, while   
$ \qbar \epsilon_{ijk}\gamma_j\gamma_k \q $ like $ J^{PC} = 1^{+-} $.   

In Fig.~\ref{fig:meson}, the ground-state masses extracted from the time-correlation fucntions
of these meson operators are plotted, 
versus the corresponding meson states in high energy experiments.
It turns out that for each theoretical state, there is an experimental counterpart 
with the same $ J^{P(C)} $, 
and the theoretical mass is in good agreement with the PDG mass, 
with the error bar (statistical and systematic combined) less than $2\%$ of its central value.
Among all states, the ground-state masses of the $\cbar \Gamma \c $ operators are in very good 
agreement with their experimental counterparts.
This is also the case for the pseudoscalar and vector mesons, as shown in the first and the third columns
in Fig.~\ref{fig:meson}. 
Moreover, it is interesting to point out that the ground-state masses of the    
scalar operators $ \ubar \d $, $ \ubar \s $, $\sbar \s$, $\ubar \c$, and $\sbar \c $ 
also agree well with 
$ f_0/\sigma_0(500) $, $ K_0^*(700) $, $f_0(980)$, $ D_0^*(2300) $, and $D_{s0}^*(2317)$ respectively.  
Similarly, the ground-state masses of the axial-vector operators $ \ubar \gamma_5 \gamma_i \d $, 
$ \ubar \gamma_5 \gamma_i \s $, $\sbar \gamma_5 \gamma_i \s$, $\ubar \gamma_5 \gamma_i \c$, 
and $\sbar \gamma_5 \gamma_i \c $ also agree well with 
$ a_1(1260) $, $ K_1(1270) $, $f_1(1285)$, $ D_{1}(2420) $, and $D_{s1}(2460)$ respectively.
Likewise, for the ground-state masses of the pseudovector meson operators 
with $ \Gamma =  \epsilon_{ijk}\gamma_j\gamma_k $ (in the last column of Fig.~\ref{fig:meson}),  
they are also compatible with 
$ h_1(1170) $, $ K_1(1400) $, $b_1(1235)$, $ D_1(2430) $, and $D_{s1}(2536)$ respectively.  

The details of all mesons in Fig.~\ref{fig:meson} will be presented in a forthcoming paper.


\begin{wrapfigure}{r}{0.7\textwidth}
\begin{center}
\includegraphics*[width=0.65\textwidth,clip=true]{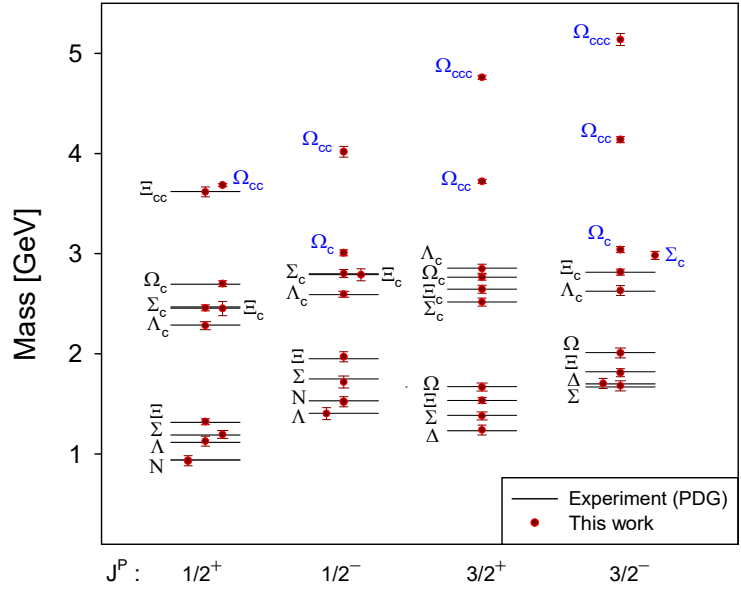}
\caption{
The ground-state masses of 3-quark baryons with $ J^P = 1/2^\pm$ and $ 3/2^\pm $,   
for all $(\u/\d, \s, \c )$ flavor combinations, versus the experimental states 
\cite{Tanabashi:2018oca}.
}
\label{fig:baryon}
\end{center}
\end{wrapfigure}

\subsection{Baryon mass spectrum}

The construction of 3-quark baryon operator and the extraction of ground-state masses from
the time-correlation function follow the prescriptions 
in our previous studies \cite{Chiu:2005zc,Chen:2017kxr}.
In Fig.~\ref{fig:baryon}, the ground-state masses of 3-quark baryons 
with $ J^P = 1/2^\pm$ and $ 3/2^\pm $ are plotted, for all $(\u/\d, \s, \c )$ flavor combinations.
They are all in good agreement with their counterparts in the experiments. 
Also, we have 9 predictions for the masses (in units of MeV) of the following charmed baryons: 
$\Sigma_c(3/2^-)[2983(34)(10)]$, $\Omega_{c}(1/2^-)[3009(32)(7)]$, $\Omega_{c}(3/2^-)[3040(26)(5)]$, 
{\small $\Omega_{cc}(1/2^+)[3624(12)(6)]$, $\Omega_{cc}(1/2^-)[4017(54)(11)]$, 
$\Omega_{cc}(3/2^+)[3721(10)(3)]$, $\Omega_{cc}(3/2^-)[4139(26)(5)]$}, 
{\small $\Omega_{ccc}(3/2^+)[4760(22)(11)]$}, {\small $\Omega_{ccc}(3/2^-)[5138(62)(15)]$}, 
where the first/second error is the statistical/systematic one. Here we identify 
$\Omega_{c}(1/2^-) $ and $\Omega_{c}(3/2^-) $ with $\Omega_c(3000)$ and $\Omega_c(3050)$ 
of the five $\Omega_c$ states observed by LHCb \cite{Aaij:2017nav} in 2017, 
and predict their $J^P$ to be $1/2^-$ and $3/2^-$ respectively. 

The details of all baryons in Fig.~\ref{fig:baryon} will be presented in a forthcoming paper.

\section{Concluding Remark}

We have generated the first gauge ensemble for $N_f = 2+1+1 $ lattice QCD 
with physical $ (\u/\d, \s, \c) $ domain-wall quarks, and determined the 
topological susceptibility (\ref{eq:chit14_MeV}) 
and extracted the ground-state hadron mass spectra in Figs.~\ref{fig:meson}-\ref{fig:baryon}. 
It is interesting to see that our theoretical hadron mass spectra are in good agreement 
with the PDG masses, plus 9 predicted states for charmed baryons.
This implies that these ground-state hadrons behave like the conventional meson/baryon  
composed of valence quark-antiquark/quark-quark-quark, interacting through the gluons
with the quantum fluctuations of $ (\u, \d, \s, \c) $ quarks in the sea. 
Note that some hadron states in Figs.~\ref{fig:meson}-\ref{fig:baryon}
are close to the threshold of strong decays, e.g.,  
$D_{s1}(2536)$ is 32 MeV above the $D^*K$ threshold,
$D_{s1}(2460)$ is 44 MeV below the $D^*K$ threshold,
and $D_{s0}^*(2317)$ is 41 MeV below the $DK$ threshold. 
This also suggests that the ground-state mesons/baryons (in Figs.~\ref{fig:meson}-\ref{fig:baryon})
couple predominantly to the quark-antiquark/3-quark interpolators, 
and only weakly (if any) to meson-meson/meson-baryon interpolators, 
not to mention diquark-antidiquark/diquark-diquark-antiquark interpolators.  


\section*{Acknowledgement}

We are grateful to Academia Sinica Grid Computing Center (ASGC)  
and National Center for High Performance Computing (NCHC) for the computer time and facilities. 
This work is supported by the Ministry of Science and Technology 
(Grant Nos.~108-2112-M-003-005, 107-2119-M-003-008).

\end{document}